\documentstyle[preprint]{ptptex}

\newcommand{\be}{\begin{equation}}
\newcommand{\ee}{\end{equation}}
\newcommand{\ba}{\begin{eqnarray}}
\newcommand{\ea}{\end{eqnarray}}
\newcommand{\nonu}{\nonumber \\}
\newcommand{\A}{&}
\newcommand{\e}{\,e}
\newcommand{\bra}[1]{\langle {#1} \vert}
\newcommand{\ket}[1]{\vert {#1} \rangle}

\preprintnumber{
STUPP--97--150 \\
January 1997
}

\title{%
The Critical Ising Model on a M\"obius Strip 
}

\author{%
Shun-ichi {\sc Yamaguchi}
}

\inst{%
Physics Department, Saitama University, Urawa 338
}

\abst{%
We study the two-dimensional critical Ising model on a M\"obius strip 
based on a duality relation between conformally invariant 
boundary conditions. By using a Majorana fermion field theory, 
we obtain explicit representations of crosscap states 
corresponding to the boundary states. 
We also discuss the duality structure of the partition functions. 
}

\begin{document}

\maketitle

The continuum limit of the two-dimensional critical Ising model 
with boundaries is one of the simplest systems realized by 
field theories with boundaries. \cite{Cardy1,BMF} \ 
The presence of boundaries gives rise to new effects 
which depend on boundary conditions. 
Simple but important boundary conditions 
for Ising spins are those of the fixed and free ones. 
These boundary conditions are conformal invariant and 
are related by duality. A model with such boundary conditions 
has been analyzed in Ref.\ \citen{Cardy1} 
using a boundary conformal field theory \cite{BCFT} 
with the central charge $c={1 \over 2}$. 
However, only a model defined on a cylinder has been considered 
to this time. 

The purpose of this paper is 
to study the critical Ising model on a M\"obius strip 
based on the duality relation 
between conformally invariant boundary conditions. 
We use a free massless Majorana fermion field theory 
as used in Ref.\ \citen{String} 
to compute loop amplitudes of superstring theories with open strings. 
We obtain explicit representations of crosscap states 
corresponding to the boundary states 
for conformally invariant boundary conditions and 
discuss the duality structure of these crosscap states. 
We also show that the periodic M\"obius partition function 
with the free boundary condition is constructed from 
those of the periodic cylinder and antiperiodic cylinder 
with free boundary conditions. 

We now consider the continuum limit of the critical Ising model 
on a M\"obius strip of width $L$ and boundary length $R$. 
We define coordinates $x,y$ as 
$0 \le x \le L$ and $0 \le y \le R$. 
When we take the coordinate $x$ to be Euclidean time, 
the system is described by 
the Hamiltonian $H_R$ for the periodic space $0 \le y \le R$. 
The partition function 
is then given by \cite{String}
\be
Z =\bra{{\rm B}} \e^{-{L \over 2} H_R} \ket{{\rm C}}\,, 
\label{Zclosed}
\ee
where $\ket{{\rm B}}$ is a boundary state placed at $x=0$ 
and $\ket{{\rm C}}$ is a crosscap state placed at $x={L \over 2}$. 
In the form (\ref{Zclosed}) the partition function is 
a function of the modular parameter 
$\tilde \tau =i{L \over R} +{1 \over 2}$. 
To represent the system concretely, we introduce free massless 
Majorana fermion fields $\psi(x,y)$ and $\overline\psi(x,y)$. 
In the picture (\ref{Zclosed}) the fields have 
standard oscillator modes $\psi_n$ and $\overline \psi_n$ for 
the periodic space $0 \le y \le R$. \cite{Ginsparg} \ 
In terms of these modes the Hamiltonian is 
\be
H_R={2 \pi \over R} \sum_{n>0} 
n \left(\psi_{-n} \psi_n +\overline \psi_{-n} \overline \psi_n \right) 
+E_0\,, 
\label{Hclosed}
\ee
where $E_0=-{\pi \over 12R}$ 
in the NS-sector $(\,n \in {\bf Z}_+ +{1 \over 2}\,)$ and 
$E_0={\pi \over 6R}$ in the R-sector $(\,n \in {\bf Z}_+\,)$. 

The explicit form of the boundary state in Eq.\ (\ref{Zclosed}) 
depends on the boundary conditions of the model. 
Here we consider fixed and free boundary conditions for Ising spins, 
which are related by duality. 
For the fixed boundary condition, there are two possibilities: 
Ising spins on the boundary are all fixed in the $s\!=\!+1$ state 
(${\rm fixed(+)}$) or in the $s\!=\!-1$ state (${\rm fixed(-)}$). 
Boundary states corresponding to the above three different boundary 
conditions were obtained in Ref.\ \citen{Cardy1}. 
In terms of the above modes they are 
\ba
\ket{{\rm B}_{\rm fixed(\pm)}} \A = \A {1 \over \sqrt 2}\,
\ket{{\rm B},-}_{\rm NS}
 \,\pm\, {1 \over \sqrt[4]{2}} \,\ket{{\rm B},-}_{\rm R} \,, 
\label{BSfixed} \\ 
\ket{ {\rm B}_{\rm free} } \A = \A \ket{{\rm B},+}_{\rm NS}
\stackrel{\rm (dual)}{=} 
{1 \over \sqrt 2} \left(\ket{ {\rm B}_{\rm fixed(+)} } 
 + \ket{ {\rm B}_{\rm fixed(-)} } \right) 
\label{BSfree}
\ea
with
\ba 
\ket{{\rm B}, \pm}_{\rm NS} \A=\A 
\e^{\pm \,i \sum_{n \geq {1 \over 2}} \psi_{-n} \overline \psi_{-n}} 
\ket{0}\,, \nonu 
\ket{{\rm B}, \pm}_{\rm R} \A=\A 
\e^{\pm \,i \sum_{n \geq 1} \psi_{-n} \overline \psi_{-n}}
\ket{\pm} \,,
\label{BSmode}
\ea
where the ground states 
in the R-sector satisfy $\psi_0 \ket{\pm}
={1 \over \sqrt 2} \e^{\pm {\pi \over 4}i} \ket{\mp}$ and 
$\overline\psi_0 \ket{\pm}
={1 \over \sqrt 2} \e^{\mp {\pi \over 4}i} \ket{\mp}$. 
The second equality of Eq.\ (\ref{BSfree}) implies that 
the free boundary condition is obtained from 
the fixed boundary conditions by duality $\psi \rightarrow \psi$, 
$\overline \psi \rightarrow -\overline \psi$. \cite{Cardy1,Cardy2} \ 
Note that the free boundary state 
defined by the duality relation (\ref{BSfree}) 
describes periodic boundary Ising spins around the boundary 
since the fixed boundary states have periodic boundary spins. 
On the other hand, the crosscap state in Eq.\ (\ref{Zclosed}) 
is generally defined by requiring the condition \cite{String}
\be
\left[T(\textstyle{L \over 2},y) 
- \overline T(\textstyle{L \over 2},y+\textstyle{R \over 2}) \right] 
\ket{\rm C} =0 \,,
\label{CScondition}
\ee
where $T(x,y)$ and $\overline T(x,y)$ are the left moving 
and right moving components of the energy momentum tensor. 
In terms of the modes, 
solutions of Eq.\ (\ref{CScondition}) are given by \cite{String}
\ba
\ket{{\rm C}, \pm}_{\rm NS} 
\A=\A \e^{\pm \,i\sum_{n \geq {1 \over 2}} \!\e^{\pi i n}\,
\psi_{-n} \overline \psi_{-n}} \ket{0} \,,\nonu
\ket{{\rm C}, \pm}_{\rm R} 
\A=\A \e^{\pm \,i \sum_{n \geq 1} \!\e^{\pi i n}\,
\psi_{-n} \overline \psi_{-n}} \ket{\pm} \,.
\label{CSmode}
\ea

To determine the crosscap states corresponding to 
the boundary states (\ref{BSfixed}) and (\ref{BSfree}) 
in Eq.\ (\ref{Zclosed}), we must evaluate these partition functions. 
They can be obtained from cylinder partition functions as follows. 
When we regard the coordinate $y$ as Euclidean time, 
the partition function (\ref{Zclosed}) has the form \cite{String}
\be
Z = {\rm tr}\,f \e^{-{R \over 2} H_L}\,, 
\label{Zopen}
\ee
where $f$ is a flip operator and 
$H_L$ is the Hamiltonian for the space $0 \le x \le L$. 
The partition function without the flip operator 
$Z^{\rm cyl}= {\rm tr} \e^{-{R \over 2} H_L}$ 
is that of the critical Ising model defined on a cylinder 
of length $L$ and circumference ${R \over 2}$. 
The M\"obius partition functions 
with fixed or free boundary conditions can be constructed from 
the cylinder partition functions with the same boundary conditions 
at the ends of the cylinder. 
There are three possible constructions: 
\be
Z^{\rm cyl\,(P)}_{\rm fixed(\pm),fixed(\pm)} 
\rightarrow Z_{\rm fixed(\pm)}\,, \qquad 
Z^{\rm cyl\,(P)}_{\rm free,free} 
\rightarrow Z_{\rm free}^{(1)}\,,\qquad 
Z^{\rm cyl\,(A)}_{\rm free,free} 
\rightarrow Z_{\rm free}^{(2)}\,, 
\label{cyltoms}
\ee
where the lower two indices in the cylinder partition functions 
are the boundary conditions at $x=0$ and $x=L$. 
${\rm (P)}$ and ${\rm (A)}$ denote 
periodic and antiperiodic boundary conditions 
of Ising spins around the cylinder, and 
${\rm (A)}$ is only allowed for the free sector. 
These two types of boundary conditions come from 
traces in the lattice transfer matrix formulation. \cite{Cardy1} \ 
We note that $Z^{\rm cyl\,(P)}_{\rm free,free}$ 
is obtained from $Z^{\rm cyl\,(P)}_{\rm fixed(\pm),fixed(\pm)} +
Z^{\rm cyl\,(P)}_{\rm fixed(\pm),fixed(\mp)}$ 
by duality (\ref{BSfree}), 
but $Z^{\rm cyl\,(P)}_{\rm fixed(\pm),fixed(\mp)}$ 
does not appear in Eq.\ (\ref{cyltoms}), and that 
$Z^{\rm cyl\,(A)}_{\rm free,free}$ cannot be expressed 
by using the boundary state (\ref{BSfree}). 

Let us now obtain explicit forms of M\"obius partition functions 
in Eq.\ (\ref{cyltoms}) following the above considerations. 
For the same spin boundary conditions at $x=0$ and $x=L$, 
the fields $\psi(x,y)$ and $\overline \psi(x,y)$ are not 
independent and are expanded as \cite{BMF}
\ba
\e^{{\pi \over 4}i} \,\psi(x,y)
\A=\A \sqrt{\pi \over L}
 \!\sum_{n \in {\bf Z}+{1 \over 2}} \! b_n 
 \e^{{\pi \over L}n(y-ix)}\,, \nonu
\e^{-{\pi \over 4}i} \,\overline\psi(x,y)
\A=\A \sqrt{\pi \over L}
 \! \sum_{n \in {\bf Z}+{1 \over 2}} \! \overline b_n \, 
 \e^{{\pi \over L}n(y+ix)}\,, 
\label{openmode}
\ea
where $\overline b_n =b_n$ for 
the fixed boundary condition and 
$\overline b_n =-b_n$ for the free boundary condition. 
The mode $b_n$ satisfies $\{b_n,b_m\}=\delta_{n+m,0}$. 
On the other hand, 
the cylinder partition functions in Eq.\ (\ref{cyltoms}) were 
obtained in terms of the $c={1 \over 2}$ Virasoro characters. 
Their explicit forms can be found in Ref.\ \citen{Cardy1}. 
Therefore, by using the mode $b_n$ representations of 
these cylinder partition functions, we can evaluate them in which 
the flip operator $f =(-1)^{\sum_{n \geq {1 \over 2}} nb_{-n}b_n}$ is 
inserted. 
In this way, we obtain 
\ba 
\A\A Z_{\rm fixed(\pm)}(\tau) 
= {1 \over 2} \left( 
Z_{\rm free}^{(1)}(\tau) +Z_{\rm free}^{(2)}(\tau) \right)\,, 
\label{Zopenfixed} \\
\A\A Z_{\rm free}^{(1)}(\tau)
= \e^{{\pi \over 48}i} \sqrt{\vartheta_3(0|\tau) \over \eta(\tau)}\,, 
\qquad 
Z_{\rm free}^{(2)}(\tau) 
= \e^{{\pi \over 48}i} \sqrt{\vartheta_4(0|\tau) \over \eta(\tau)}\,, 
\label{Zopenf}
\ea
where we have introduced 
Jacobi theta functions $\vartheta_i(z|\tau)$ and 
the Dedekind eta function $\eta(\tau)$ with modular parameter 
$\tau =i{R \over 4L} +{1 \over 2}$. 
We use the notation of Ref.\ \citen{Ginsparg}. 

To compare the above partition functions 
with those in the form (\ref{Zclosed}), 
we change the variable $\tau$ to $\tilde\tau$ in
Eqs.\ (\ref{Zopenfixed}) and (\ref{Zopenf}). 
The transformation properties of $\vartheta$ and $\eta$ functions 
under the modular transformation 
$\tau \rightarrow \tilde\tau={\tau -1 \over 2 \tau -1}$ 
can be obtained by using the Poisson resummation formula. 
The results are 
\ba
\vartheta_3 (0|\tau) \A=\A 
\e^{{\pi \over 2}i}\,(-2 \tilde\tau +1)^{1 \over 2}\, 
\vartheta_4 (0|\tilde\tau)\,,\nonu 
\vartheta_4 (0|\tau) \A=\A
(-2 \tilde\tau +1)^{1 \over 2}\, 
\vartheta_3 (0|\tilde\tau)\,,\nonu
\eta(\tau) \A=\A
\e^{{\pi \over 4}i}\,(-2 \tilde\tau +1)^{1 \over 2}\, 
\eta(\tilde\tau)\,. 
\label{transf}
\ea
Substituting Eq.\ (\ref{transf}) into Eqs.\ (\ref{Zopenf}) and 
(\ref{Zopenfixed}), we obtain the partition functions 
as functions of $\tilde\tau$. 
On the other hand, 
by using Eqs.\ (\ref{Hclosed})$\sim$(\ref{BSfree}) and (\ref{CSmode}), 
these partition functions can be uniquely expressed as 
\ba 
Z_{\rm fixed(\pm)} (\tilde\tau) \A=\A 
\bra{ {\rm B}_{\rm fixed(\pm)} } \e^{-{L \over 2} H_R} \,
\ket{ {\rm C}_{(\rm fixed)} } \,, 
\nonu
Z_{\rm free}^{(1)} (\tilde\tau) \A=\A
\bra{ {\rm B}_{\rm free} }
\e^{-{L \over 2} H_R} \,
\ket{ {\rm C}_{({\rm free})}^{(1)} } \,, 
\nonu
Z_{\rm free}^{(2)} (\tilde\tau) \A=\A
\bra{ {\rm B}_{\rm free} }
\e^{-{L \over 2} H_R} \,
\ket{ {\rm C}_{({\rm free})}^{(2)} } \,, 
\label{Zclosedff}
\ea
where 
\ba
\ket{{\rm C}_{\rm (fixed)}} \A=\A 
 {1 \over \sqrt 2} \left( \e^{+{\pi \over 8}i} \ket{{\rm C},+}_{\rm NS} 
+ \e^{-{\pi \over 8}i} \ket{{\rm C},-}_{\rm NS} \right) \,,
\nonu
\ket{{\rm C}_{\rm (free)}^{(1)}} \A=\A
 \e^{+{\pi \over 8}i} \ket{{\rm C},-}_{\rm NS} \,,
\nonu
\ket{{\rm C}_{\rm (free)}^{(2)}} \A=\A
 \e^{-{\pi \over 8}i} \ket{{\rm C},+}_{\rm NS} \,.
\label{Cff}
\ea
Thus we have obtained the crosscap states 
corresponding to the boundary states (\ref{BSfixed}) and (\ref{BSfree}). 

Let us discuss the above results. 
We first find that 
the crosscap state $\ket{{\rm C}_{\rm (fixed)}}$ 
constructed by only the NS-sectors corresponds to 
two different fixed boundary states 
$\ket{{\rm B}_{\rm fixed(+)}}$ and $\ket{{\rm B}_{\rm fixed(-)}}$. 
This implies that the R-sector in the fixed boundary states 
(\ref{BSfixed}) makes no contribution to the partition 
function $Z_{{\rm fixed(\pm)}}$. 
Next, we remark that $Z_{\rm free}^{(1)}$ and $Z_{\rm free}^{(2)}$ 
are described by 
the same boundary state $\ket{{\rm B}_{\rm free}}$, and therefore 
not only $Z_{\rm free}^{(1)}$ but also $Z_{\rm free}^{(2)}$ 
have periodic boundary Ising spins around the boundary. 
This can be understood as follows. 
If we define a crosscap state $\ket{{\rm C}_{({\rm free})}}$ by 
\be
\ket{{\rm C}_{({\rm free})}} ={1 \over \sqrt 2} 
\left(\ket{{\rm C}_{({\rm free})}^{(1)}} +
\ket{{\rm C}_{({\rm free})}^{(2)}}\right)\,,
\label{Cfree}
\ee
it satisfies $\ket{{\rm C}_{({\rm free})}} \stackrel{\rm (dual)}{=}
\ket{{\rm C}_{({\rm fixed})}}$. 
Then, from Eqs.\ (\ref{BSfree}), (\ref{Zclosedff}) and 
(\ref{Cfree}) we obtain a duality relation 
between fixed and free boundary conditions 
in the form (\ref{Zclosed}) as 
\be
\bra{ {\rm B}_{\rm free} }
\e^{-{L \over 2} H_R} \,\ket{ {\rm C}_{({\rm free})} }
\stackrel{\rm (dual)}{=}
{1 \over \sqrt 2} \left[\left(
\bra{{\rm B}_{\rm fixed(+)}}
+ \bra{{\rm B}_{\rm fixed(-)}}\right) 
\! \e^{-{L \over 2} H_R} \,\ket{{\rm C}_{({\rm fixed})}}\right]\,. 
\label{dualclosed}
\ee 
Therefore, from Eqs.\ (\ref{Cfree}) and (\ref{dualclosed}) 
we finally obtain the following identification 
in Eq.\ (\ref{Zopenfixed}) 
\be
Z_{\rm free} \equiv 
{1 \over 2} \left( Z_{\rm free}^{(1)} +Z_{\rm free}^{(2)} \right) \,. 
\label{Zfree}
\ee
We see that $Z_{\rm free}$ defined by Eq.\ (\ref{Zfree}) 
is just the M\"obius partition function 
with the free boundary condition. 
It is identical to $Z_{\rm fixed(\pm)}$ 
in Eq.\ (\ref{Zopenfixed}), showing the duality relation 
between fixed and free boundary conditions 
corresponding to the form (\ref{dualclosed}). 

\vspace{0.5\baselineskip}


The author would like to thank Y. Tanii 
for careful reading of this manuscript 
and making many valuable comments. 
He is also grateful to K. Hida for useful discussions. 




\begin{thebibliography}{99}
%
\bibitem{Cardy1} J.L. Cardy, 
Nucl.\ Phys.\ {\bf B275} (1986), 200; {\bf B324} (1989), 581. 
%
\bibitem{BMF} S. Ghoshal and A. Zamolodchikov, 
Int.\ J.\ Mod.\ Phys.\ {\bf A9} (1994), 3841. 
\\
R. Chatterjee, 
Mod.\ Phys.\ Lett.\ {\bf A10} (1995), 973. 
%
\bibitem{BCFT} J.L. Cardy, 
Nucl.\ Phys.\ {\bf B240} (1984), 514. 
%
\bibitem{String} C.G. Callan, C. Lovelace, C.R. Nappi and S.A. Yost, 
Nucl.\ Phys.\ {\bf B293} (1987), 83. 
\\
J. Polchinski and Y. Cai, 
Nucl.\ Phys.\ {\bf B294} (1988), 91. 
%
\bibitem{Ginsparg} P. Ginsparg, 
in {\it Fields,\ Strings and Critical Phenomena, 
Les Houches XLIX}, 
ed.\ E. Brezin and J. Zinn-Justin (North Holland, 1989). 
%
\bibitem{Cardy2} J.L. Cardy, 
in {\it Fields,\ Strings and Critical Phenomena, 
Les Houches XLIX}, 
ed.\ E. Brezin and J. Zinn-Justin (North Holland, 1989). 
%
\end{thebibliography}
\end{document}